\documentstyle[epsfig]{mn}

\begin{document}

\title[Be/X-ray transient V0332+53]{The Be/X-ray Transient V0332+53:
Evidence for a tilt between the orbit and the equatorial plane?}

\author[I.~Negueruela et al.]
{I.~Negueruela$^{1,2}$\thanks{email:ignacio@tocai.sdc.asi.it},
P.~Roche$^{3}$, J.~Fabregat$^{4}$
and M.~J.~Coe$^{5}$\\
$^{1}$Astrophysics Research Institute, Liverpool John Moores University,
Byrom St., Liverpool, L3 3AF, U.K.\\
$^{2}${\em SAX Science Data Center}, ASI,
c/o Nuova Telespazio, via Corcolle 19, I00131 Rome, Italy\\
$^{3}$Astronomy Centre, CPES, University of Sussex, Falmer, Brighton, 
BN1 9QH, U.K.\\
$^{4}$Departamento de Astronom\'{\i}a y Astrof\'{\i}sica, Universidad de 
Valencia, 46100 Burjassot, Valencia, Spain\\
$^{5}$Physics and Astronomy Department, Southampton University,
Southampton, SO17 1BJ, U.K.}

\date{Received    ; 
Accepted     }

\maketitle 

\begin{abstract}
We present optical and infrared observations of
BQ Cam, the optical counterpart to the Be/X-ray transient system V0332+53. 
BQ Cam is shown to be an O8\,--\,9Ve star, which places V0332+53 at a 
distance of $\sim 7$ kpc. H$\alpha$ spectroscopy and infrared photometry 
are used to discuss the evolution of the circumstellar envelope.  
Due to the low inclination of the system,
parameters are strongly constrained. We find strong evidence for a tilt
of the orbital plane with respect to the circumstellar disc (pressumably
on the equatorial plane).
Even though the periastron distance is only $\approx 10
R_{*}$, during the present quiescent state the circumstellar 
disc does not extend to the distance of periastron passage. Under
these conditions, X-ray emission is effectively prevented by centrifugal
inhibition of accretion. The circumstellar disc is shown to be optically
dense at optical and infrared wavelengths, which together with its small
size, is taken as an indication of tidal truncation.

\end{abstract}


\section{Introduction}

The hard X-ray transient V0332+53 (X\,0331+53) was first detected by 
the {\em Vela 5B} satellite during a bright outburst in 1973 (Terrel \& 
Priedhorsky 1984). It was rediscovered ten years later by {\em Tenma} 
during a series of smaller outbursts (Tanaka et al. 1983). 
{\em EXOSAT} observations 
were used to determine that the source pulsates with a period of 4.4 s 
(Stella et al. 1985). Doppler shifts in pulse arriving times indicate
that the pulsar is in a 34.25-d binary orbit with an eccentricity 
$e =0.31$ (Stella et al. 1985). The 
optical counterpart was identified by Honeycutt \& Schlegel (1985) as the 
heavily-reddened early-type star BQ Cam. This object was observed to display 
highly variable H$\alpha$ emission (Corbet, Charles \& van der Klis 1986, 
and references therein) and infrared excess (Coe et al. 1987). These 
characteristics are typical of a Be/X-ray binary. 

In this subclass of Massive X-ray Binaries, 
the X-ray emission is believed to be due to accretion of matter from 
a Be star by a compact companion (see White, Nagase \& Parmar 1995; 
Negueruela 1998). The name 
``Be star'' is used as a general term describing an early-type 
non-supergiant star, which at some time has shown emission in the Balmer 
series lines (Slettebak 1988, for a review). Both the emission
lines and the characteristic strong infrared excess when compared to
normal stars of the same spectral types are attributed to the presence
of a circumstellar disc. Most
Be/X-ray binaries have relatively eccentric orbits and the neutron star 
companion is normally far away from the disc surrounding the Be star.
Due to their different geometries and the varying 
physical conditions in the circumstellar disc, Be/X-ray binaries can present 
very different states of X-ray activity (Stella, White \& Rosner 1986). 
In 
quiescence, they display persistent low-luminosity ($L_{{\rm x}} \la 10^{36}$ 
erg s$^{-1}$) X-ray emission or no detectable emission at all. 
Occasionally, they show 
series of periodical (Type I) X-ray outbursts ($L_{{\rm x}} 
\approx 10^{36} - 10^{37}$ erg s$^{-1}$), at the time of periastron 
passage of the neutron star (e.g., A\,0535+26, Motch et al. 1991). More 
rarely, they undergo giant (Type II) X-ray outbursts ($L_{{\rm x}} 
\ga 10^{37}$ erg s$^{-1}$), which do not show clear orbital modulation. Some 
systems only display persistent emission, but most of them show outbursts and 
are termed Be/X-ray transients.

\begin{figure*}
\begin{center}
\epsfig{file=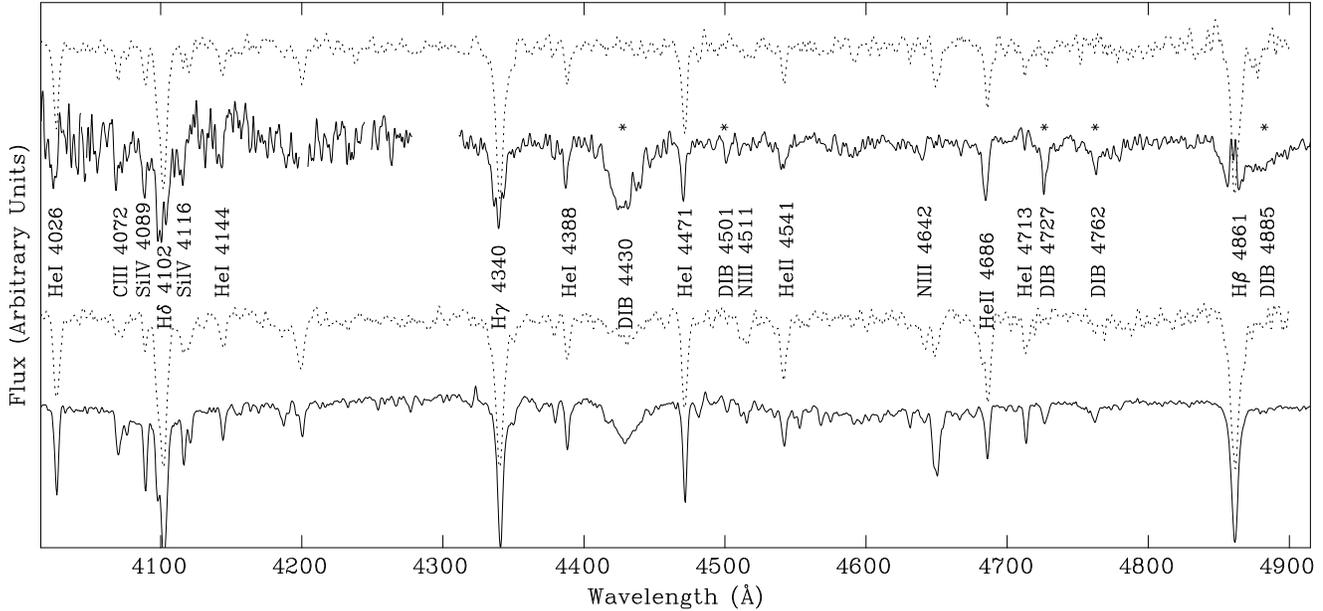, width=18.6 cm, 
bbllx=105pt, bblly=200pt, bburx=525pt, bbury=390pt, clip=}
 \end{center}
\caption{Composite spectrum of BQ Cam in the classification region. 
The short-wavelength end
was taken on August 14, 1997, using ISIS on the WHT equipped with the 
R1200R grating and the Loral 1 CCD. The long-wavelength stretch was taken 
on November 17, 1997, with the same instrument equipped with the R1200B 
grating and the EEV\#10 camera. The comparison spectra (dashed) are those of
the O9III standard $\iota$ Ori (top) and the O8V standard HD 48279, both
taken from the digital atlas of Walborn \& Fitzpatrick (1990). Also displayed
(bottom) is the spectrum of the O9IIe star HD 333452, from Steele et al.
(1999). All spectra have been smoothed with a $\sigma = 0.6$\AA\ Gaussian 
function for display. Note the strong interstellar lines in the spectrum
of BQ Cam, marked with a `*'.} 
\label{fig:blue} 
\end{figure*}

Like most other Be/X-ray transients,
V0332+53 has shown both types of outbursts. The 1973 outburst lasted $\sim 100$
days and peaked at $\sim 1.6$ Crab near July 10. It was clearly a Type II 
outburst, even though Whitlock (1989) found an underlying orbital modulation 
when the main trend was removed. On the other hand, the three weak outbursts 
separated by the orbital period observed in 1983 are Type I outbursts. During 
these outbursts, the pulsed fraction was small ($10-15$\%) and the temporal 
behaviour was dominated by random rapid fluctuations (Makishima et al. 1990a). 
The spectrum was fitted with a power-law modified by cyclotron absorption.
 Unger et al. (1992) found that 
the pulse profile varied between a double-peaked and a single-peaked structure.
The equivalent hydrogen column density remained relatively constant at $\sim 
1\times10^{22}$ atom cm$^{-2}$. A prominent absorption feature at 28.5 keV, if 
attributed to electron cyclotron resonance, implies a magnetic field at 
the surface of the neutron star of $2.5\times10^{12}$G 
(Makishima et al. 1990b).

A new outburst was discovered by $Ginga$ in September-October 1989. The 
source remained very bright for more than two weeks, indicating that this 
was a Type II outburst. A quasi-periodic oscillation, possibly 
implying the presence of an accretion disc, was detected 
(Takeshima et al. 1994). V0332+53 has not been detected
by the BATSE experiment since the {\em CGRO} satellite started operations 
in April 1991  (Bildsten et al. 1997). It is not detected with 
any significance by the All-Sky Monitor (ASM) on board {\em RXTE} either, 
according to the quick-look results provided by the {\em RXTE}/ASM 
team. Stella et al. (1986) interpret the lack of quiescence emission as 
an effect of centrifugal inhibition of accretion.

\section{Observations}

We present data obtained as a part of the Southampton/Valencia/SAAO
long-term monitoring campaign of Be/X-ray binaries (see Reig et al. 1997a), 
consisting of optical spectroscopy, infrared and optical 
broad-band photometry of BQ Cam, the optical counterpart to V0332+53.

\subsection{Blue optical spectroscopy}

The source was observed on August 14, 1997 using the 4.2-m William Herschel 
Telescope (WHT), located at the Observatorio del Roque de los Muchachos, 
La Palma, 
Spain. The blue arm was equipped with the Loral1 CCD and the R1200R grating, 
which gives a nominal dispersion of $\sim 0.25$ \AA/pixel.
A second observation was taken on November 14, 1997. On this occasion the blue 
arm was equipped with the R1200B grating and the EEV\#10 CCD, giving a nominal
dispersion of $\sim 0.22$ \AA/pixel over $\sim 900$ \AA. A composite spectrum
is shown in Fig. \ref{fig:blue}. The signal-to-noise
ratio (SNR) of the August spectrum is relatively low. The Loral camera 
introduces several artifacts in the range $\lambda$ 4150\,--\,4250 \AA,
where the spectrum could well be dominated by noise. A very strong spurious
feature was present at the wavelength where He\,{\sc ii} $\lambda$ 4200 \AA\ 
should be found. 
Between $\lambda$ 4320 \AA\ and $\lambda$ 4500 \AA, where both 
spectra overlap, all the features look very similar and only
the higher quality November spectrum is shown.

\subsection{Red optical spectroscopy}

We have monitored the source since 1990, using the 2.5-m
Isaac Newton Telescope (INT) and 4.2-m WHT, both located at the 
Observatorio del Roque de 
los Muchachos, and the 1.5-m telescope at Palomar Mountain (PAL).
A log of observations, together with some parameters measured, is 
presented in Table \ref{tab:halpha}. The Palomar spectra have
relatively low SNR and the error in EW, arising due to the difficulty
of determining the continuum, is $\sim 15\%$. The line shapes are
also difficult to establish. In contrast, the INT and WHT spectra have all 
relatively high resolution and errors in the EW of H$\alpha$ are $\sim 5\%$.
The double-peaked structure of H$\alpha$ is only clearly visible in the 
spectra with dispersions of 0.4 \AA/pixel or better.

All the data have been reduced using the {\em Starlink}
software packages {\sc ccdpack} (Draper 1998) and {\sc figaro} 
(Shortridge et al. 1997) and
analysed using {\sc figaro} and {\sc dipso} (Howarth et al. 1997). 

H$\alpha$ spectra normally show He\,{\sc i} $\lambda$6678 \AA\ as a very
weak emission feature when the SNR is sufficiently large for it to be
separated from the noise. An analysis of H$\alpha$ variability in BQ Cam
during 1990\,--\,1991 has been presented in Negueruela et al. (1998). Many 
of the spectra whose parameters are listed in Table \ref{tab:halpha} are 
shown in their Fig. 8, and are therefore not reproduced here. 
Negueruela et al. (1998) found that during
1990\,--\,1991, the H$\alpha$ line presented V/R variability with
a quasi-period of $\sim 1$ year, but this variability stopped late
in 1991.

\begin{table}
\caption{Details of the H$\alpha$ spectroscopy. The FWHM of H$\alpha$ has 
been calculated by fitting a single Gaussian to the line profile and should
therefore be taken only as an approximation. In the higher resolution 
spectra, however, this procedure
gives values consistent with direct measurement to within 5\%.}
\begin{center}
\begin{tabular}{lcccc}
\hline 
Date of       & Telescope & Dispersion & EW of & FWHM\\
Observation(s)&           & (\AA/pixel) & H$\alpha$ (\AA) & (km s$^{-1}$)\\
\hline
&&&\\
Jan 28, 1990 & INT & $\sim 0.2$ & $-$4.5 & 250 \\
Sep 02, 1990 & INT & $\sim 0.8$ & $-$5.0 & 250 \\
Nov 14, 1990 & INT & $\sim 0.8$ & $-$5.7 & 260 \\
Dec 27, 1990 & INT & $\sim 0.4$ & $-$6.8 & 260 \\
Jan 27, 1991 & INT & $\sim 0.8$ & $-$6.5 & 240\\
Aug 28, 1991 & INT & $\sim 0.4$ & $-$4.6 & 280\\
Dec 14, 1991 & INT & $\sim 0.4$ & $-$4.8 & 270\\
Aug 18, 1992 & PAL & $\sim 0.9$ & $-$3.1 & 230 \\
Nov 13, 1992 & PAL & $\sim 0.9$ & $-$3.8 & 290 \\
Dec 05, 1993 & PAL & $\sim 0.9$ & $-$6.8 & 280 \\
Aug 14, 1997 & WHT & $\sim 0.4$ & $-$4.2 & 280\\
Nov 14, 1997 & WHT & $\sim 0.4$ & $-$3.9 & 270\\
\hline
\end{tabular}
\end{center}
\label{tab:halpha}
\end{table}

\subsection{Optical Photometry}

Optical photometry of the source was obtained on November 11, 1997, using the 
1-m Jakobus Kapteyn Telescope (JKT) at the Observatorio del Roque de los 
Muchachos, La Palma, Spain. The telescope was equipped with the Tek4 CCD 
and the Harris filter set. Conditions were photometric. Instrumental 
magnitudes were extracted 
through synthetic aperture routines contained in the {\sc iraf} package, 
and transformed to the Johnson/Cousins system through calibrations 
derived from observations of a number of Landolt (1992) standard stars taken 
on the same night.  
The values measured are $U=17.74\pm0.10$, $B=17.29\pm0.02$, $V=15.73\pm0.02$, 
$R=14.69\pm0.02$ and $I= 13.27\pm0.08$. Errors in the $U$ and $I$ bands are 
dominated by calibration uncertainties (mostly in the colour correction 
equations). The smaller errors in $B,V$ and $R$ are dominated by measurement
errors.

\subsection{Infrared Photometry}

\begin{figure*}
\begin{center}
    \leavevmode
\epsfig{file=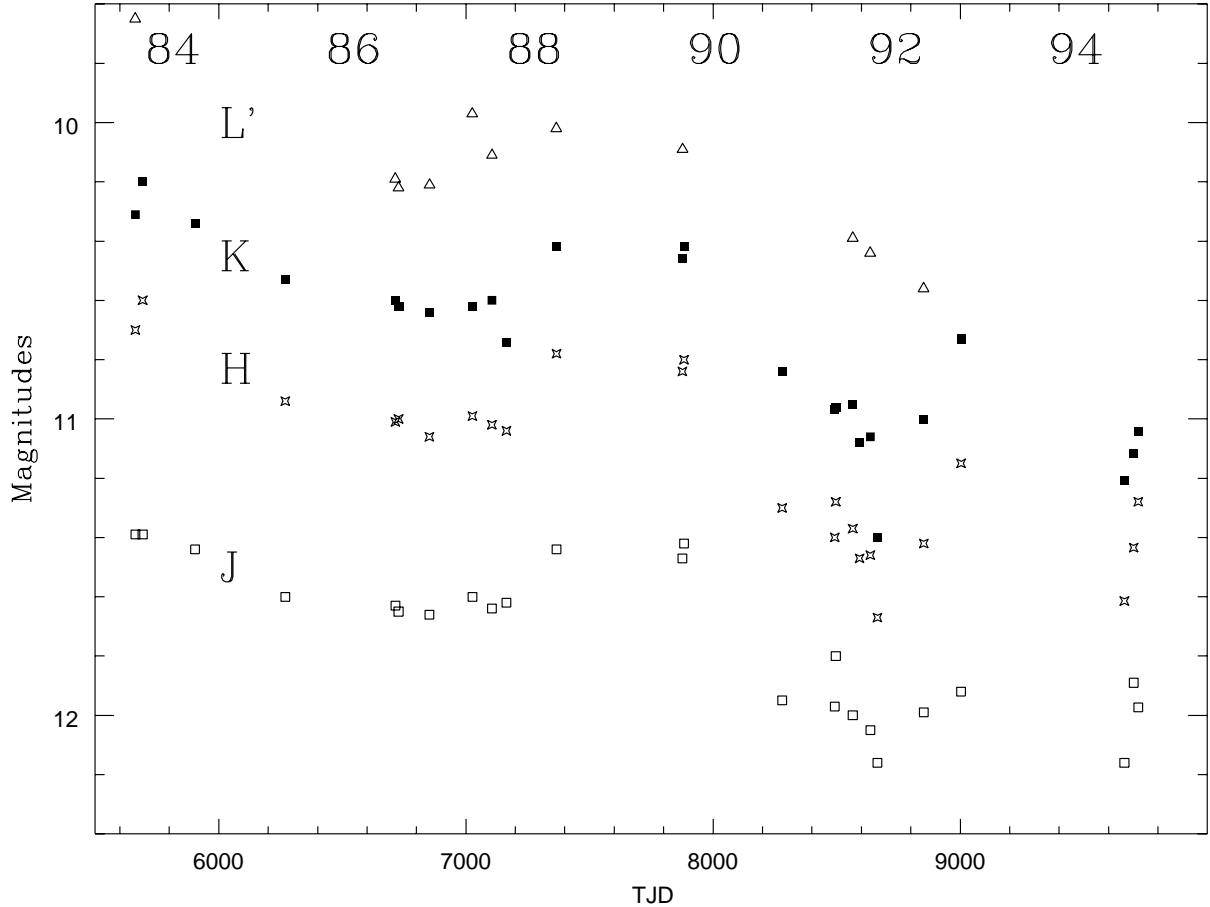, width=16cm, bbllx=30pt,
bblly=25pt, bburx=525pt, bbury=400pt, clip=}
 \end{center}
\caption{Infrared lightcurves of BQ Cam during 1983\,--\,1995. $J$ magnitudes
are represented by open squares, $H$ by stars, $K$ by filled squares and
$L'$ by triangles. Error bars have been removed for clarity (see Table 
\ref{tab:infra} for details).}
\label{fig:irlc}
\end{figure*}

Infrared observations of BQ Cam are listed in Table \ref{tab:infra}.
They have been obtained with  
the Continuously Variable Filter (CVF) on the 1.5-m Carlos S\'{a}nchez
Telescope (TCS) at the Teide Observatory, Tenerife, Spain, and the UKT9
detector at the 3.9-m UK Infrared Telescope (UKIRT) at the Mauna Kea
observatory on Hawaii. The December
1994 observation was taken with IRCAM mounted on UKIRT and 540-s total
exposure in each filter. Data for the 
period 1983\,--\,1986 have already been reported in Coe et al. (1987). The
observations presented here extend for a further ten years. The
long-term lightcurve is shown in Fig. \ref{fig:irlc}. 

\begin{table*}
\begin{center}
\begin{tabular}{lccccc}
\hline
Date &TJD & J & H & K & L$^{\prime}$ \\
&&&&&\\
\hline
&&&&&\\
1983 Nov 23 &45662(U)&11.39$\pm$0.03& 10.70$\pm$0.03& 10.31$\pm$0.03&9.6$\pm$0.05\\    
1983 Dec 23 &45692(U)&11.39$\pm$0.03& 10.60$\pm$0.03& 10.20$\pm$0.03&$-$\\            
1984 Jul 22 &45904(U)&11.44$\pm$0.03&       $-$     & 10.34$\pm$0.03&$-$\\ 	
1985 Jul 22 &46269(U)&11.60$\pm$0.03& 10.94$\pm$0.03& 10.53$\pm$0.03&$-$\\
1986 Oct 10 &46714(U)&11.63$\pm$0.03& 11.01$\pm$0.03& 10.60$\pm$0.03&10.19$\pm$0.05\\
1986 Oct 23 &46727(U)&11.65$\pm$0.03& 11.00$\pm$0.03& 10.62$\pm$0.03&10.22$\pm$0.05\\ 
1987 Feb 25 &46852(U)&11.66$\pm$0.03& 11.06$\pm$0.03& 10.64$\pm$0.03&10.21$\pm$0.05\\ 
1987 Aug 18 &47026(U)&11.60$\pm$0.03& 10.99$\pm$0.03& 10.62$\pm$0.03&9.97$\pm$0.09\\ 
1987 Nov 05 &47105(U)&11.64$\pm$0.03& 11.02$\pm$0.03& 10.60$\pm$0.03&10.11$\pm$0.11\\ 
1988 Jan 03 &47164(T)&11.62$\pm$0.04& 11.04$\pm$0.04& 10.74$\pm$0.04&$-$\\
1988 Jul 23 &47366(U)&11.44$\pm$0.02& 10.78$\pm$0.02& 10.42$\pm$0.03&10.02$\pm$0.08\\ 
1989 Dec 15 &47876(U)&11.47$\pm$0.01& 10.84$\pm$0.01& 10.46$\pm$0.01&10.09$\pm$0.08\\ 
1989 Dec 21 &47882(U)&11.42$\pm$0.01& 10.80$\pm$0.01& 10.42$\pm$0.01&$-$\\
1991 Jan 22 &48279(T)&11.95$\pm$0.06& 11.30$\pm$0.03& 10.84$\pm$0.03&$-$\\
1991 Aug 23 &48492(T)&11.97$\pm$0.04& 11.40$\pm$0.03& 10.97$\pm$0.03&$-$\\
1991 Aug 25 &48494(T)&12.08$\pm$0.06& 11.36$\pm$0.03& 11.01$\pm$0.04&$-$\\
1991 Aug 27 &48496(T)&11.80$\pm$0.05& 11.28$\pm$0.04& 10.96$\pm$0.04&$-$\\
1991 Nov 04 &48565(U)&12.00$\pm$0.02& 11.37$\pm$0.01& 10.95$\pm$0.02&10.39$\pm$0.06\\
1991 Dec 01 &48592(T)&12.2 $\pm$0.2 & 11.47$\pm$0.07& 11.08$\pm$0.06&$-$\\
1992 Jan 14 &48636(U)&12.05$\pm$0.01& 11.46$\pm$0.01& 11.06$\pm$0.01&10.44$\pm$0.04\\
1992 Feb 12 &48665(T)&12.16$\pm$0.04& 11.67$\pm$0.05& 11.40$\pm$0.06&$-$\\
1992 Aug 17 &48852(U)&11.99$\pm$0.01& 11.42$\pm$0.02& 11.00$\pm$0.01&10.56$\pm$0.09\\
1993 Jan 15 &49003(T)&11.92$\pm$0.07& 11.15$\pm$0.04& 10.73$\pm$0.03&$-$\\
1994 Nov 07 &49664(T)&12.16$\pm$0.06& 11.61$\pm$0.03& 11.21$\pm$0.05&$-$\\
1994 Dec 14 &49701(U)& 11.89$\pm$0.02 & 11.43$\pm$0.02& 11.12$\pm$0.02&$-$\\
1995 Jan 02 &49720(T)&11.97$\pm$0.14& 11.28$\pm$0.05& 11.04$\pm$0.06&$-$\\
\hline
\hline
\end{tabular}
\end{center}
 \caption{Observational details of the IR photometry. Observations marked T 
are from the TCS, while those marked U are from UKIRT. The first five 
observations are taken from Coe et al. (1987).}
 \label{tab:infra}
\end{table*}

\section{Results}

\subsection{Spectral classification}

The spectrum of BQ Cam in the classification region is displayed in Fig. 
\ref{fig:blue}. We can see that H$\beta$ is in emission, with a clear
double-peaked structure (see also Fig. \ref{fig:highres}), almost to the
continuum level. The He\,{\sc i} lines at $\lambda\lambda$ 4713 \& 5016 \AA\
also show weak double-peaked emission, just above the continuum level. In 
contrast,  H$\gamma$ only shows two very weak emission components on the 
wings of the absorption line. 

The very strong diffuse interstellar lines are consistent with the high
reddening of the object. The strength of the He\,{\sc ii} lines clearly
identifies the object as an O-type star. This is confirmed by the strong
Si\,{\sc iv} lines. 
Overall the spectrum is very similar to that of LS 437, the optical 
counterpart to 3A\,0726$-$260 (Negueruela et al. 1996). 

In  Fig. \ref{fig:blue}, the spectrum of BQ Cam is compared to that of
two MK standards from the digital atlas of Walborn \& Fitzpatrick (1990),
$\iota$ Orionis (O9III) and HD 48279 (O8V). Also shown is a high SNR
spectrum of the Be star HD 333452 (O9IIe) from Steele, Negueruela
\& Clark (1999). Points to be noted are:
\begin{itemize}
\item The H\,{\sc i} and He\,{\sc i} lines are weaker in BQ Cam than in 
any standard, due to the presence of circumstellar emission.
\item BQ Cam must be later than O7, since  He\,{\sc i} $\lambda$4471 \AA 
 $>>$ He\,{\sc ii} $\lambda$4541 \AA.
\item The presence of He\,{\sc i} $\lambda \lambda$ 4026, 4144 \AA\,
the strength of the Si\,{\sc iv} doublet, C\,{\sc iii} $\lambda$4072 \AA\ and 
He\,{\sc i} $\lambda$4388 \AA\ all point to BQ Cam being no earlier
than O8.
\item The main luminosity criterion, namely, the strength of the 
Si\,{\sc iv} $\lambda \lambda$ 4089, 4116 \AA\ doublet is difficult to 
judge in BQ
Cam, because He\,{\sc i} $\lambda$4121 \AA\ is not present (presumably 
filled-in by emission) and the quality of the spectrum is low at that end.
\item The Mg\,{\sc ii} $\lambda$4481 \AA\ line is visible at this resolution
in both evolved O9 stars, but it is absent in BQ Cam, indicating a lower
luminosity class. Similarly, \hbox{N\,{\sc iii}} $\lambda\lambda$ 4511-4515 
\AA\
is much weaker in BQ Cam than in the giants. This line is also stronger
in HD 48279, but Walborn \& Fitzpatrick (1990) warn that this object
is N-enhanced.
\item The N\,{\sc iii} lines are only seen in absorption in the 
O8\,--\,O9 range in main-sequence stars (Walborn \& Fitzpatrick 1990).
N\,{\sc iii} $\lambda\lambda$ 4379, 4642 \AA\ and possibly 
N\,{\sc iii} $\lambda$4511 \AA\ are visible on the spectrum of BQ Cam.
\item The equivalent width (EW) of  
He\,{\sc ii} $\lambda$4686 \AA\ in BQ Cam (0.6 \AA) is inside the 
range typical for O8\,--\,O9 stars (Conti \& Alschuler 1971).
\item The O\,{\sc ii} $\lambda$4367 \AA\ line, which can be seen in the
spectra of the two O9 stars is also visible in the spectra of the O9V
standards 10 Lac and HD 93028 at this resolution, hinting that BQ Cam
could be earlier.
\item The complete absence of C\,{\sc iii} $\lambda$4650 \AA\ in BQ Cam is
surprising. In main sequence stars, it is clearly visible as early as O8
(Walborn \& Fitzpatrick 1990). This line is also absent in LS 437. There
is no reason to think that BQ Cam is carbon deficient, since 
C\,{\sc iii} $\lambda$4072 \AA\ is present and there is no sign of 
N-enhancement. One possible explanation is that the inner regions of 
the circumstellar discs of Oe stars are hot enough to produce C\,{\sc iii} 
emission, which would be filling in the $\lambda$4650 \AA\ line. 
\end{itemize}

From all the above, it is clear that BQ Cam is an unevolved star in the
O8\,--\,9 range. An O8.5V classification is very likely, but, since the
presence of emission affects the main classification criteria, we prefer
to be cautious and give an spectral type O8-9Ve.
 
\subsection{Distance}

In order to obtain an estimate of the distance to the system, it is 
necessary to determine the interstellar absorption in its direction. This
is complicated because Be stars present circumstellar reddening, due to
the infrared continuum emission. The
calculation by Kodaira et al. (1985) of $A_{V}=7.4$ is a gross overestimate,
since they took the infrared magnitudes measured in 1983 to be those
intrinsic to the star, while our data show that the disc emission accounted
for at least $\sim 0.8$ mag in $J$. Our $UBVRI$ photometry was taken at
a time when the circumstellar disc was certainly small (see Sections
\ref{sec:par} and \ref{sec:discuss}). The small EW of H$\alpha$ at the
time of the observations (see Table \ref{tab:highres}) should be 
accompanied by a small circumstellar reddening (see Dachs,
Engels \& Kiehling 1988).

Since the intrinsic colour of a late O-type star is $(B-V)_{0}=-0.31$ 
for luminosity classes III\,--\,V (Schmidt-Kaler 1982),
the measured $(B-V)=1.56\pm0.03$ implies
$E(B-V)=1.87$. This value is almost identical to the $E(B-V)=1.88\pm 0.1$
deduced from the strength of different interstellar bands by Corbet et al.
(1986), though the photometric determination is more reliable. It is also
compatible within the errors with the value of 
$N_{{\rm H}}$ derived by Unger
et al. (1992) from $EXOSAT$ X-ray data taken during the 1983 Type I 
outbursts, which implies $E(B-V) \approx 1.7$ (Bohlin, Savage \& Drake 
1978). Since a very
small amount of circumstellar reddening could be present, the derived
distance should be taken as a lower limit. Zorec \& Briot (1991) 
and Fabregat \& Torrej\'{o}n (1998) have noted that Be stars are on 
average 0.3 mag brighter than main-sequence objects (due to the added 
luminosity of the circumstellar disc). Though the disc surrounding BQ Cam 
must be small, some contribution to the absolute luminosity should be
expected. However, we will use the  absolute magnitude of a normal O8.5V
star $M_{V}=-4.5$ (Vacca, Garmany \& Shull 1996), once again taking into 
account that the distance calculated will be a lower limit.

Using the standard reddening of $R=3.1$, we derive $d=7.6$ kpc. However,
using Schmidt-Kaler's (1982) expression for the reddening to O stars, we find 
$R=3.3$, which gives a distance of $d=6.3$ kpc. Given the uncertainty
in $R$ and taking into account the above considerations, we will accept the 
distance $d=6$ kpc as a lower limit (unless the reddening in that direction
is exceptionally strong). An estimate of the different factors mentioned
above, would indicate a range $6<d<9$ kpc for the distance to BQ Cam.

We note that for an O9III star, $M_{V}=-5.5$ (Vacca et al. 1996) and the
implied distance is $d \ga 10$ kpc, which would place the object well 
outside the galactic disc. This is taken as confirmation of the main-sequence
classification for BQ Cam.

\subsection{System parameters}
\label{sec:par}

The very low mass function of V0332+53, $f(M) = 0.10\pm0.03$ (Stella et al. 
1985) indicates that
the orbit of the neutron star is seen under a very small inclination angle.
Assuming a lower limit for the mass of an O-type companion  
$M_{*}\ga20M_{\sun}$ 
and the standard mass for a neutron star $M_{{\rm x}}=1.44\,M_{\sun}$, 
an inclination angle $i\la 10^{\circ}.3\pm0^{\circ}.9$ is obtained. 
Waters et al. (1989) have argued that the orbital
plane of Be/X-ray binaries with close orbits should not be very inclined 
with respect to the equatorial plane in which the circumstellar disc is
supposed to form. 

We have used our high-resolution spectra from 
November 14, 1997 to measure the parameters of several emission and
absorption lines, which are listed in Table \ref{tab:highres}, in order
to estimate the rotational velocity of BQ Cam. This is particularly
difficult given the presence of a circumstellar component. We have selected
the two strongest He lines in the blue.
He\,{\sc ii} $\lambda$4686 \AA\ is not likely to be contaminated by any 
emission component from the disc, but can be affected by non-LTE effects.
On the other hand, He\,{\sc i} $\lambda$4471 \AA\ is likely to be
affected by circumstellar emission, which will reduce its FWHM. Therefore
any estimation of $v \sin i$ based on this line should be taken as 
a lower limit.

From Buscombe's (1969) approximation, the measured FWHMs corrected for
instrumental broadening imply  $v\sin i= 160\, {\rm km}\,{\rm s}^{-1}$ 
from the He\,{\sc ii} line and $v\sin i= 130\, {\rm km}\,{\rm s}^{-1}$ 
from the He\,{\sc i} line. 
Similarly, using the correlation between FWHM of He\,{\sc i} $\lambda$4471 
\AA\ and $v \sin i$ from Slettebak et al. (1975), we obtain 
$v\sin i= 135\, {\rm km}\,{\rm s}^{-1}$. The two values derived from the 
He\,{\sc i} $\lambda$4471 \AA\ line are very similar and set a lower limit 
for $ v\sin i$.

A very different way of estimating the rotational velocity is by
using the mean relation between FWHM
of the H$\alpha$ emission line and  $v\sin i$ for Be stars
from Hanuschik, Kozok \& Kaiser (1988). Since the FWHM of H$\alpha$
in BQ Cam had not changed significantly during 6 years, we deduce that
the disc is dynamically stable and the correlation can be trusted to a
relatively high degree. We note that the scatter in the correlation is
due to the inclusion of measurements for stars with dynamically unstable
discs.

Using

\[
\log \frac{{\rm FWHM}}{2(v\sin i)} = -0.2 \log W_{\alpha} +0.11
\]
we obtain $v\sin i= 140\, {\rm km}\,{\rm s}^{-1}$. Similarly, using the mean 
relation between peak 
separation of H$\alpha$ and $v\sin i$ for Be stars (Hanuschik et al. 1988), 
\[
\log \frac{\Delta v_{\rm peak}}{2(v\sin i)} = -0.4 \log W_{\alpha} -0.1
\]
we obtain $v\sin i= 170\, {\rm km}\,{\rm s}^{-1}$

\begin{figure}
\begin{center}
    \leavevmode
\epsfig{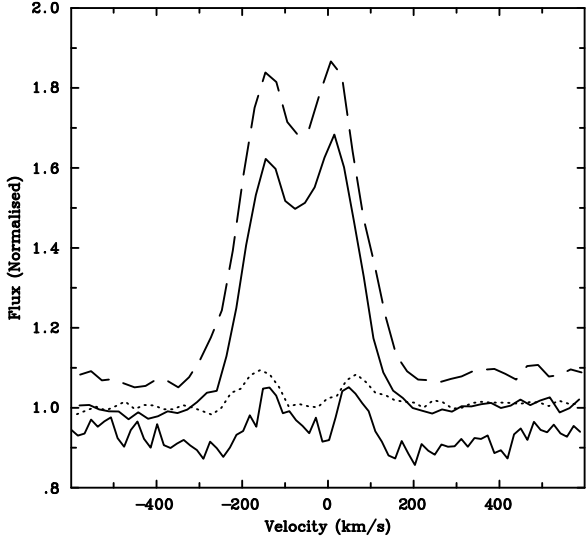}
 \end{center}
\caption{Emission lines in BQ Cam observed on November 14, 1997 with 
the WHT equipped with the the R1200B grating and the EEV\#10 CCD on the
blue arm and the R1200R grating and the Tek5 CCD on the red arm. From
bottom to top,  H$\beta$, He\,{\sc i} $\lambda$6678\AA\ (dotted line)
and H$\alpha$. 
All the lines have been normalised to the continuum level (note that
H$\beta$ emission is inside the wings of the photospheric absorption 
and therefore below the continuum level). For the line parameters, see
Table \ref{tab:highres}. Also plotted is the H$\alpha$ profile from
December 12, 1991, displaced by 0.1 units of the continuum. Note the 
similarity in shape, intensity and FWHM.}
\label{fig:highres}
\end{figure}

All the above estimates provide similar values. The estimates based on
He\,{\sc i} $\lambda$4471 \AA\ yield a lower limit for $v \sin i \ga
130\, {\rm km}\,{\rm s}^{-1}$. Averaging the four estimates gives a
value of $\sim 150\, {\rm km}\,{\rm s}^{-1}$. The errors associated with 
this estimation are formally large. However, the shape and FWHM of 
H$\alpha$ are not compatible with a very low $v \sin i$ ($\la 100\, 
{\rm km}\,{\rm s}^{-1}$), while a value approaching
$v \sin i \approx 200\, {\rm km}\,{\rm s}^{-1}$ does not seem compatible
with the comparatively small width (both at the base and at half-maximum)
of H$\alpha$ when a large population of Be stars are considered (see,
for example, Hanuschik et al. 1996).

Since all Be stars are believed to be fast rotators, this value of
$v \sin i$ confirms that the star is seen under a small inclination
angle. We note, however, that in order to show a $v \sin i \approx 150
\, {\rm km}\,{\rm s}^{-1}$ with an inclination angle 
$i=10^{\circ}$, the rotational velocity of the star should be $v \sim 900$ 
km s$^{-1}$, well above the break-up velocity of a late O-type star ($\la
600$ km s$^{-1}$). Assuming an upper limit for the rotational velocity of
$v = 0.8 v_{{\rm break}} \la 480$ km s$^{-1}$, still gives $i \ga 
19^{\circ}$. An orbital inclination of $i= 19^{\circ}$, would imply
an enormously undermassive primary with $M_{*}=5 M_{\sun}$ (the errors
in the orbital parameters allow up to $M_{*}=7 M_{\sun}$ within 2-$\sigma$). 
Even if we
assume that the O star is rotating at break-up velocity, $i=15^{\circ}$
implies $M_{*} = 8M_{\sun}$, still undermassive by a factor $> 2$. 
We note the uncertainty in our estimate of $v \sin i$, but
the  constraints $v < 600$ km s$^{-1}$ and $i\la10^{\circ}$ can only be 
compatible with $v \sin i \la 100$ km s$^{-1}$. This is not only far away 
from our estimate of $v \sin i$, but also very difficult to reconcile
with the FWHM and shape of H$\alpha$ (see Hummel 1994, Hanuschik et al. 1996). 

On the other hand, we have no strong reasons to
expect a very undermassive optical star. The calculations by
Vanbeveren \& De Loore (1994) show that, under certain circumstances, 
mass transfer in massive binaries can lead to the formation of
overluminous post-main-sequence stars with compact companions 
(e.g., Vela X-1). It is not clear how noticeable this effect will be 
as long as the star which has received mass
remains in the main sequence, and whether this will affect 
$T_{{\rm eff}}$ (and therefore, spectral class). Gies et al. (1998)
have found evidence suggesting that the optical component of the
Be + sdO binary $\phi$ Per is moderately undermassive, but no evidence 
exists for any main-sequence component of an X-ray binary being
undermassive by a factor $>2$. As a consequence, we 
believe that the discrepancy in the two values for $\sin i$
strongly suggests that the orbital plane is not exactly aligned with the 
equatorial plane of the Be star, even though the difference may be 
small ($\sim 10^{\circ}$).

Hummel (1994) has shown that, for inclination angles $i \la 30^{\circ}$,
the profile of emission lines from Be stars is dominated by the flank 
inflections generated by non-coherent scattering,
giving rise to what is known as the wine-bottle shape. Wine-bottle
shapes are found for Be stars with $v \sin i \la 250$ km s$^{-1}$ and
Hanuschik et al. (1996) estimate that flank inflections are visible 
for inclinations up to $i \sim 60^{\circ}$. However, the 1997 H$\alpha$ 
profiles of BQ Cam have no sign of flank inflections
(see Fig. \ref{fig:highres}). Since it is not reasonable to 
suppose that $i > 60^{\circ}$ for BQ Cam, we interpret the absence of 
flank inflections as proof that the envelope of BQ Cam is small and 
the optical depth in the vertical direction is not large enough to 
produce the wine-bottle profile typical of non-coherent scattering. 
Therefore the peak separation of emission lines in November 1997
will reflect the actual extent of the disc. 

\subsection{Disc evolution}

Iye \& Kodaira (1985) and Corbet et al. (1986) describe radial-velocity 
changes in the H$\alpha$ emission line and investigate their possible
connection with the orbital period. Our H$\alpha$ spectroscopy shows
that these changes were also present during 1990\,--\,1991, but 
Negueruela et al. (1998) have shown that these velocity variations 
can be explained by quasi-cyclic V/R variability with a quasi-period 
$\sim 1$ year. Similar cyclic variability is seen in many other 
Be/X-ray binaries 
(Negueruela et al. 1998). It is noteworthy that the system was displaying
V/R variability in 1983\,--\,1984 when it showed a short span of X-ray
activity  and again in 1990, immediately after the 1989 Type II outbursts.
The possibility of a causal connection
between V/R variability and X-ray activity has been discussed in Negueruela
et al. (1998). 

\begin{table}
\begin{center}
\begin{tabular}{lccc}
\hline 
Line & EW (\AA)& FWHM & Peak separation\\
&& (km s$^{-1}$)&(km s$^{-1}$) \\
\hline
&&&\\
H$\alpha$ & $-$3.9 & 270 & 155\\
H$\beta$ & $\sim 0.8$ & $-$ & 200\\
He\,{\sc i} $\lambda$6678\AA\ & $-$0.4 & $-$ & 230\\
He\,{\sc ii} $\lambda$4686\AA\ & 0.6 & 270 & $-$\\
He\,{\sc i} $\lambda$4471\AA\ & 0.6 & 220 & $-$\\ 
\hline  
\end{tabular}
\end{center}
\caption{Details of several lines in the spectrum of BQ Cam, observed 
on November 14, 1997, with the WHT. Emission lines are shown 
in Fig. \ref{fig:highres}. The EW of H$\beta$ is only approximate, due
to the contamination by the interstellar band at $\lambda$ 4885 \AA. 
Note that He\,{\sc ii} $\lambda$ 4686\AA\ and He\,{\sc i} $\lambda$4471\AA\
are photospheric absorption lines, though the latter could be affected 
by a circumstellar emission component.}
\label{tab:highres}
\end{table}

The infrared lightcurves (see Fig. \ref{fig:irlc}) of BQ Cam show a general
fading trend, only interrupted by a brief brightening in 1988\,--\,1989. 
The brightest magnitudes observed are those from late 1983, when the
source was active in the X-rays, coinciding with the highest  EWs of 
H$\alpha$ reported ($-8$ and $-10$ \AA; Kodaira et al. 1985; Stocke et al.
1985). The decline of the strength of H$\alpha$ (Iye \& Kodaira 1985) was
accompanied by the fading of infrared magnitudes. Corbet et al. (1986) 
and Coe et al. (1987) interpret the decline as being due to the dispersion 
of the circumstellar disc of the Be star. 

The infrared magnitudes remained stable during 1985\,--\,1987, but 
brightened again in 1988, reaching values similar to those of 1983.
After the type II outburst in late 1989, the infrared magnitudes faded
to a deeper minimum, where they have remained until 1995, though showing
considerable short-time variability. There does not
seem to be any corresponding systematic change in H$\alpha$ EW (see Table
\ref{tab:halpha}). The variability in H$\alpha$ during 1990\,--\,1991 can
be associated with the V/R cycle seen at the time (Negueruela et al. 1998).
Given the similarity between the 
relatively high-resolution spectra of 28 August 1991, 14 December 1991, 
14 August 1997 and 14 November 1997, it seems unlikely that any 
significant V/R variability has been present after 1991.

It is noteworthy that, while the infrared colours have experimented large
fluctuations ($\sim 0.8$ mag in $J$ and $\sim 1.2$ in $K$), the associated
colours have remained much more stable (with $(J-K) \sim 0.8\,-\,1.2$).
There is no clear correlation between the brightness and the colours. If, 
for instance, we compare the $J$ magnitude with  $(J-K)$, we see that
the faintest observations can be either very blue (TJD 48665) or
relatively red (TJD 48494). The brightest observations are on average 
relatively red, but not more than some faint points.

Using the correlations of Rieke \& Lebofsky (1985), from the observed 
$A_{V}=R \times E(B-V)=6.17$ we deduce $A_{J}=1.74$ and $A_{K}=0.69$, 
implying an interstellar reddening $E(J-K)=1.05$. Using the fainter and
bluer infrared observations (TJD 48665), we find $J_{0}=J-A_{J}=10.42$
and $K_{0}=K-A_{K}=10.71$, implying $(J-K)_{0}=-0.29\pm0.07$, which is
roughly compatible with Wegner's (1994) average value of $(J-K)_{0}=-0.18$ 
for O9V stars, if we 
consider the errors in his value and in the standard relations used.
The measured $E(J-K)=(J-K)-(J-K)_{0}=0.94$ is, within the errors,
compatible with the value for interstellar reddening found above and 
implies that no circumstellar reddening was present. This
corresponds to a state in which the disc is optically thin at all infrared
wavelengths and very little infrared emission is produced (see Dougherty
et al. 1994). Brighter magnitudes with a similarly blue colour (as in TJD
49701) must represent a state in which the disc is producing a significant
amount of infrared emission, but still remains optically thin, giving rise
to no circumstellar reddening -- a condition which could be associated
with a very small disc (Dougherty et al. 1994). Very faint magnitudes
with redder colours (as in TJD 49664) represent states in which little 
emission is present, but the disc has become (partially) optically thick 
in $K$, which can be associated with a larger disc or with a change
in the density gradient. When the disc is very bright (as in 1983 or 1988),
it is always relatively red. The circumstellar emission is very intense
(contributing $\sim 1$ mag), but the disc is optically thick in all
wavelengths, and the circumstellar reddening remains constant at a low
value, presumably because the disc is still small (Dougherty et al. 1994).

\section{Discussion}
\label{sec:discuss}

We have shown that the optical counterpart to V0332+53 is an O8--9Ve star,
further skewing the spectral distribution of Be/X-ray binary transients 
towards earlier spectral types (see Negueruela 1998).
Our distance estimate is higher than those of previous authors, who 
observed brighter magnitudes due to larger disc contamination. 
Honeycutt \& Schlegel (1985) report $B=17.04\pm0.06$ and $(B-V)=
1.62\pm0.06$ in two separate observations taken on November 27-28, 1983, 
and February 21, 1984, while our observations show magnitudes fainter 
by $\sim 0.3$ mag, though the values for
$(B-V)$ are compatible within the observational errors. Normally observed 
photometric variability of classical Be stars is typically $\la 
0.2\,{\rm mag}$. Therefore BQ Cam does not show variations as large as
V635 Cas, the optical counterpart to 4U\,0115+634
(Negueruela et al. 1997), but it is more variable than most classical Be 
stars.  
At a distance of $\sim 7$ kpc, the maximum X-ray luminosity of 
V0332+53 observed by $Vela$ 5$B$ is $L_{{\rm x}} \ga 10^{38}$ 
erg s$^{-1}$, close to the Eddington luminosity for a neutron star.
At this distance,
BQ Cam is not likely to be part of the Perseus arm, but rather of
an outer galactic arm (see Kimeswenger \& Weinberger 1989).

With an inclination angle $i \la 10^{\circ}$, the orbital solution
for V0332+53
implies $a_{{\rm x}} \ga 8.5\times10^{10}$ m. For a companion mass 
$M_{*}\ga 20M_{\sun}$, this implies a periastron distance $a_{{\rm per}} \ga 
6.3\times10^{10}$ m $\approx 10 R_{*}$, where the radius of the optical
star is assumed to be $R_{*} = 9R_{\sun}$ (Vacca et al. 1996). Using our
value for $v \sin i$ and Huang's (1972) law
\[
\frac{R_{{\rm d}}}{R_{*}} = \left( \frac{2v\sin i}{\Delta v_{{\rm peak}}} 
\right)^{2}
\]
where $v_{{\rm peak}}$ is the separation between the peaks of an emission 
line and  Keplerian rotation of the envelope is assumed (which gives upper
limits), we obtain outer emission radii $R_{{\rm d}} = 4.0, 2.5$ and $1.8 
R_{*}$ for
H$\alpha$, H$\beta$ and  He\,{\sc i} $\lambda$6678\AA\  in November 1997. 
This clearly shows that the neutron star does not come close to the dense 
regions 
of the circumstellar disc in the present situation. The neutron star is
only reached by the low-density outer envelope and accretion is 
centrifugally inhibited (Stella et al. 1986). The reduced size of the
circumstellar disc strongly points at the possibility of disc truncation
by the neutron star, an idea advanced by Okazaki (1998) and supported
by the results of Reig, Fabregat \& Coe (1997b). 

This truncation would explain why we observe instances of a small disc
which is optically thick at all wavelengths, implying a very high 
density. We note that V635 Cas shows large brightness variations with
little change in the associated colours (Negueruela et al. 1997), a 
behaviour that could be associated with a very optically thick disc 
(Dougherty et al. 1994). This variability extends to the $B$ band, which 
can change by at least $0.6\,{\rm mag}$. It seems likely then that the
disc in BQ Cam is not so optically thick as that in V635 Cas at
optical wavelengths, but can be very optically thick in the infrared.
Roche et al. (1999) find that optical and infrared observations
of the Be/X-ray transient Cep X-4 are also best explained with a 
truncated dense disc.

We have established that the orbit of the neutron star is likely to be 
inclined with respect to the equatorial plane of the Be star (in which 
the disc is suppose to lie while it is dynamically stable). This is in 
contradiction with the general argument presented by Waters et al. (1989),
but it is in not unexpected. We note that the Be + neutron star system
PSR B1259$-$63, which is believed to be a representative of the class
of systems which will evolve into Be/X-ray binaries, is likely to have
a tilted orbit with respect to the equatorial plane of the Be star (see 
Ball et al. 1999 and references therein), and 
the B + neutron star system PSR J0045$-$7319, which must have formed
in a way analogous to Be/X-ray binaries, has been shown to have a rotation 
axis misaligned with the orbit (Kaspi et al. 1996). The relevance of this
misalignment to the formation and evolution of binary systems containing
neutron stars has been discussed by van den Heuvel \& van Paradijs (1997)
and Iben \& Tutukov (1998)

Corbet \& Peele (1997) have suggested a possible 34.5-d period for the
Be/X-ray binary 3A\,0726$-$260. If this period was to be confirmed, the 
comparison between V0332+53 and  3A\,0726$-$260 would be most interesting,
since the two systems would have neutron stars orbiting Oe stars of almost
identical spectral types with extremely similar orbital periods. In contrast,
the X-ray activity of these two systems is completely different. V0332+53
is a transient, which spends most time in quiescence and shows very bright
outbursts, while 3A\,0726$-$260 seems to be a persistent low-luminosity 
source with small outbursts.
The difference in pulse periods (4.4-s against 103.2-s) could be reflecting
the very different behaviour of accreted material in magnetic fields of
very different intensity.
However, we note that the quiescent luminosity of  3A\,0726$-$260 is 
almost identical to that of A\,0535+262, which has almost the same spin 
period, but a much broader orbit. This, together with the fact that the 
source lies nowhere close to the $P_{{\rm orb}}/P_{{\rm s}}$ relationship 
for Be/X-ray binaries, casts some
doubt on the orbital period until it can be confirmed using orbital Doppler
shift in the arrival time of pulses.

\section{Conclusions}

We have presented long-term photometry and spectroscopy of the optical
component of the Be/X-ray binary V0332+53, which indicate that it is an
O8--9Ve star at a distance of $\sim 7$ kpc. We find evidence for a tilt
of the orbital plane with respect to the equatorial plane. The lack of 
recent X-ray activity is explained by the fact that the dense regions of 
the circumstellar disc around the Oe star do not reach the orbit of 
the neutron star. The low 
inclination of the orbit allows us to determine a periastron
distance $a_{{\rm per}} \ga 10 R_{*}$, while measurements from our
high-resolution spectroscopy of 
emission lines set the outer radius of the H$\alpha$ emitting region at 
$R_{{\rm d}} \la 4 R_{*}$. Under these conditions, centrifugal inhibition
of accretion effectively prevents any X-ray emission.

\section*{Acknowledgements}

We would like to thank the UK PATT and the Spanish CAT panel for supporting 
our long-term monitoring campaign. We are very grateful to the INT and WHT
service programmes for obtaining most optical observations. The
1.5-m TCS is operated by the Instituto de Astrof\'{\i}sica de Canarias at the
Teide Observatory, Tenerife. The WHT and INT are operated on the island of La 
Palma by the Royal Greenwich Observatory in the Spanish Observatorio del 
Roque de Los Muchachos of the Instituto de Astrof\'{\i}sica de Canarias. 
The 1.5-m telescope at Mount Palomar is jointly owned by the California
Institute of Technology and the Carnegie Institute of Washington. We
are very grateful to all astronomers who have taken part in
observations for this campaign, G. Capilla, D. Chakrabarty, J.~S.~Clark,
C. Everall, J. Grunsfeld, A.~J.~Norton, H. Quaintrell, P. Reig, 
A. Reynolds, J.~B.~Stevens, J.~M.~Torrej\'{o}n and S.~J.~Unger. James
Stevens obtained and reduced the optical photometry of BQ Cam. This 
research has made use of the Simbad database,
operated at CDS, Strasbourg, France. The data reduction was
carried out using the Southampton University  and Liverpool 
John Moores University {\em Starlink} nodes, which are funded by PPARC. 
At Liverpool IN was funded by PPARC, while now he holds an ESA external
fellowship. An anonymous referee is thanked by helpful
remarks which helped to improve the paper.

\end{document}